\documentclass[]{aastex6}

\shortauthors{Bower et al.}
\shorttitle{V830 Tau}
\begin{document}

\newcommand\degd{\ifmmode^{\circ}\!\!\!.\,\else$^{\circ}\!\!\!.\,$\fi}
\newcommand{\etal}{{\it et al.\ }}
\newcommand{\uv}{(u,v)}
\newcommand{\rdm}{{\rm\ rad\ m^{-2}}}
\newcommand{\msuny}{{\rm\ M_{\sun}\ y^{-1}}}
\newcommand{\mylesssim}{\stackrel{\scriptstyle <}{\scriptstyle \sim}}
\newcommand{\lsim}{\stackrel{\scriptstyle <}{\scriptstyle \sim}}
\newcommand{\gsim}{\stackrel{\scriptstyle >}{\scriptstyle \sim}}
\newcommand{\sci}{Science}
\newcommand{\sgr}{PSR J1745-2900}
\newcommand{\sgra}{Sgr~A*}
\newcommand{\kms}{\ensuremath{{\rm km\,s}^{-1}}}
\newcommand{\masy}{\ensuremath{{\rm mas\,yr}^{-1}}}

\newcommand{\vtau}{V830 Tau}

\def\kbar{{\mathchar'26\mkern-9mu k}}
\def\totd{{\mathrm{d}}}


\title{Variable Radio Emission from the Young Stellar Host of a Hot Jupiter}

\author{Geoffrey C.\ Bower}
\affil{Academia Sinica Institute of Astronomy and Astrophysics, 645 N. A'ohoku Place, Hilo, HI 96720, USA}
\email{ gbower@asiaa.sinica.edu.tw}
\author{Laurent Loinard}
\affil{Instituto de Radioastronom\'ia y Astrof\'isica, Universidad Nacional Aut\'onoma de Mexico, Morelia 58089, Mexico}
\affil{Max Planck Institut f\"ur Radioastronomie, Auf dem H\"ugel 69, D-53121 Bonn, Germany}
\author{Sergio Dzib}
\affil{Max Planck Institut f\"ur Radioastronomie, Auf dem H\"ugel 69, D-53121 Bonn, Germany}
\author{Phillip A.B. Galli}
\affil{Univ. Grenoble Alpes, IPAG, 38000, Grenoble, France}
\affil{CNRS, IPAG, F-38000 Grenoble, France}
\affil{Instituto de Astronomia, Geof\'isica e Ci\^encias Atmosf\'ericas, Universidade de Sa\~o Paulo, Rua do Mata\~o 1226, Cidade Universit\'aria,  05508-900, S\~ao Paulo - SP, Brazil}
\author{Gisela N. Ortiz-Le\'on}
\affil{Instituto de Radioastronom\'ia y Astrof\'isica, Universidad Nacional Aut\'onoma de Mexico, Morelia 58089, Mexico}
\author{Claire Moutou}
\affil{Canada France Hawaii Telescope Corporation, CNRS, 65-1238 Mamalahoa Hwy, Kamuela, HI 96743, USA}
\affil{Aix Marseille Universit\'e, CNRS, LAM (Laboratoire d'Astrophysique de Marseille) UMR 7326, 13388, Marseille, France }
\and
\author{Jean-Francois Donati}
\affil{Universit\'e de Toulouse, UPS-OMP, IRAP, 14 avenue E. Belin, Toulouse, F-31400, France}
\affil{CNRS, IRAP/UMR 5277, Toulouse, 14 avenue E. Belin, F-31400, France}

\begin{abstract}
We report the discovery of variable radio emission associated with
the T Tauri star, \vtau, which was recently shown to host a hot
Jupiter companion. Very Large Array observations at a frequency of 6 GHz reveal a detection
on 01 May 2011 with a flux density $919 \pm 26\ \mu$Jy, along with 
non-detections in two other epochs at $<66$ and $<150\ \mu$Jy.  Additionally,
Very Long Baseline Array observations include one detection 
and one non-detection at comparable sensitivity, demonstrating that the emission is nonthermal
in origin.  The emission is consistent with the gyro-synchrotron
or synchrotron mechanism from a region with a magnetic field
$\gsim 30$ G, and is likely driven by an energetic event such
as magnetic reconnection that accelerated electrons.  With the limited
data we have, we are not able
to place any constraint on the relationship between the radio emission and
the rotational or orbital properties of \vtau.  This is the first 
detection of radio emission from a non-degenerate star known to host an
exoplanet.
\end{abstract}


\section{Introduction}

High energy processes, including those that produce radio emission, 
are an important diagnostic of stellar magnetospheres
and the influence of those magnetospheres on accretion,
planet formation, and even habitability \citep{1999ARA&A..37..363F,2002ApJ...574..258F,2002ARA&A..40..217G,2012IAUS..286..405A,2015ApJ...809...79O}.   
Of the thousands of known exoplanets, hot Jupiters 
represent the best opportunity to study the interaction between
star and exoplanet due to the close proximity, which creates 
amplified stellar wind flux and potentially an enhanced
exoplanet magnetosphere that
could stimulate strong radio emission such as that seen
in Jovian decametric bursts \citep{2015aska.confE.120Z}.


Aside from the first discovered exoplanets found orbiting a pulsar
\citep{1992Natur.355..145W}, however, none of the known exoplanets or their stellar hosts have been detected 
at radio wavelengths.  The absence of stellar host
detection is partly due to selection effects
in the optimal exoplanet search methods that select against the most active stars, which
are the stars most likely to produce radio emission \citep{2002ARA&A..40..217G}.  Primarily,
however, it is a result of the very low luminosity of thermal emission
from ordinary stars at radio
wavelengths.  Exoplanet emission is even more challenging because of the likely weaker magnetic
field strengths in these bodies.  
While Jupiter is the brightest low frequency radio source in the solar system,
when moved to a distance of 10 pc, its flux density falls below that of the faintest radio source
ever detected.  Nevertheless, hot Jupiters have been considered promising targets because
of their proximity to the host star and possibly higher magnetic field strengths in these
objects.  Numerous searches at long radio wavelengths have been carried out 
without clear evidence of detection of either star or exoplanet
\citep[e.g.,][]{2003ASPC..294..151F,2004IAUS..213...73F,2007ApJ...668.1182L,2010AJ....140.1929L,2009A&A...500L..51L,2013ApJ...762...34H,2014A&A...562A.108S,2015EGUGA..1714693W}.

The recent discovery of a hot Jupiter companion to \vtau\ has raised the possibility of
study of an exoplanet associated with a young, active system \citep{2015MNRAS.453.3706D,2016Natur.534..662D}.
\vtau\
is a non-accreting solar-mass T Tauri star located at 150 pc \citep{2009ApJ...698..242T}. Its age is estimated from pre-main-sequence evolutionary tracks 
to be $\sim$ 2 Myr  \citep{2000A&A...358..593S}.
\vtau\ has been observed in the framework of the \added{Magnetic Topologies of Young Stars \& the Survival of close-in massive Exoplanets(MaTYSSE)}\footnote{http://matysse.irap.omp.eu} program, which focuses on the role of the magnetic field in the formation of stars and planets using high-resolution optical spectropolarimetry. The star was first observed with  the \added{Echelle SpectroPolarimetric Device 
for the Observation of Stars} \citep[ESPaDOnS;][]{2003ASPC..307...41D} at the Canada-France-Hawaii Telescope (CFHT) on
top of Maunakea in December 2014. 

The large-scale magnetic topology at the stellar surface was modeled from a time series of total intensity and circular polarization profiles, while unpolarized
intensity velocity profiles showed a hint of residual periodic signal after the large velocity jitter from the rotation-modulated spot pattern has been removed \citep{2015MNRAS.453.3706D}. A more intensive observing campaign was repeated during the following season, from November to December 2015 using CFHT/ESPaDOnS, its twin Narval on Telescope Bernard Lyot (TBL), and \added{Gemini Remote Access to CFHT ESPaDOnS Spectrograph} \citep[GRACES;][]{2014SPIE.9151E..47C}. These new data confirmed the presence of a hot Jupiter in a 4.93d orbit around this 2 Myr T Tauri star \citep{2016Natur.534..662D}. This discovery is important for the implication of formation and migration timescales of giant planets in close orbits around solar-type stars and for the potential impact of young hot Jupiters on the early architecture of planetary systems.   

The rotation properties of the host star were also accurately derived from the Doppler imaging data, including a constraint on the differential rotation. \vtau\ has a rotation period of 2.741d 
\citep[in agreement with][]{2013AstL...39..251G}
and a differential rotation rate between the equator and pole of 0.0172 $\pm$ 0.0014 radians per day. Concerning the large-scale structure of the surface magnetic field, Zeeman Doppler Imaging allowed derivation of a field with topology that is dominated by a dipole having an average unsigned flux of 350 G. The reconstructed brightness map of V830 Tau shows that about 12\% of the surface is covered by cool spots and bright plages over a wide range of latitudes and that the stellar spin axis is tilted by 55$^{\circ}$ with respect to the line of sight  \citep{2015MNRAS.453.3706D,2016Natur.534..662D}.

In this paper, we describe analysis of archival Karl G. Jansky Very Large Array (VLA) and
Very Long Baseline Array (VLBA) observations of \vtau.  
In section 2, we present the observations and demonstrate a detection of variable radio emission
from \vtau.  In section 3, we discuss these observations in the context of known emission
for T Tauri stars and the properties of \vtau, in particular.  We also consider optimal
methods for direct or indirect detection of the exoplanet through radio techniques.

\section{Observations and Results}

\subsection{Very Large Array}

Very Large Array observations of \vtau\ were conducted on three separate
epochs in Spring 2011.
Statistics on variability and spectral index from these observations were summarized in \citet{2015ApJ...801...91D} as part of a large survey.  
Total integration
on \vtau\ was approximately 4 minutes per epoch.  The VLA was in its
B configuration for these observations, leading to synthesized beam
sizes of $\sim 1$ arcsecond.  All observations were obtained in
a dual-polarization wide band continuum mode providing a total of 2 GHz of bandwidth, with
1 GHz of bandwidth centered at 4.5 GHz and 1 GHz of bandwidth centered
at 7.5 GHz.  Absolute flux calibration was provided with a short 
observation of 3C 147.  Amplitude and phase gain calibration was
provided with observations of the compact source J0403+2600.  Data
were calibrated, flagged, and imaged using standard methods in the CASA package.

\begin{figure}
\includegraphics[width=\textwidth]{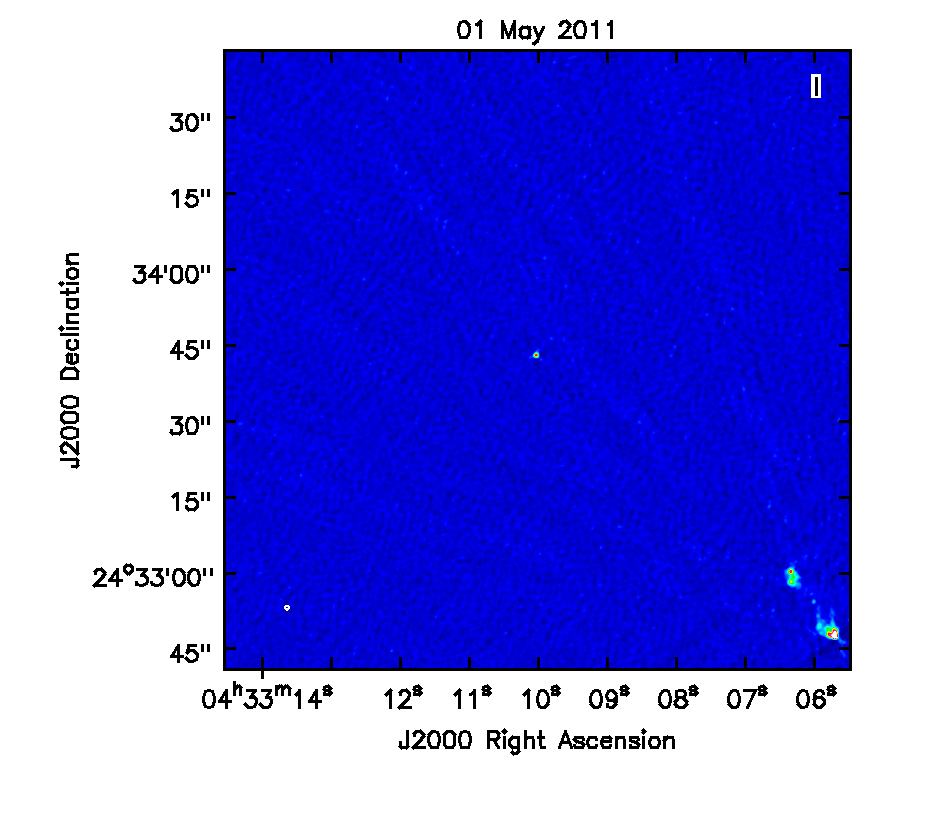}
\caption{V830 Tau on 01 May 2011 in a wide field image.  \vtau\ is at
the center of the image.  A bright, likely extragalactic source is located
to the southeast of \vtau.  The synthesized beam is shown in the lower left.
\label{fig:widefield}
}
\end{figure}

\begin{figure}[htb]
\includegraphics[width=0.33\textwidth]{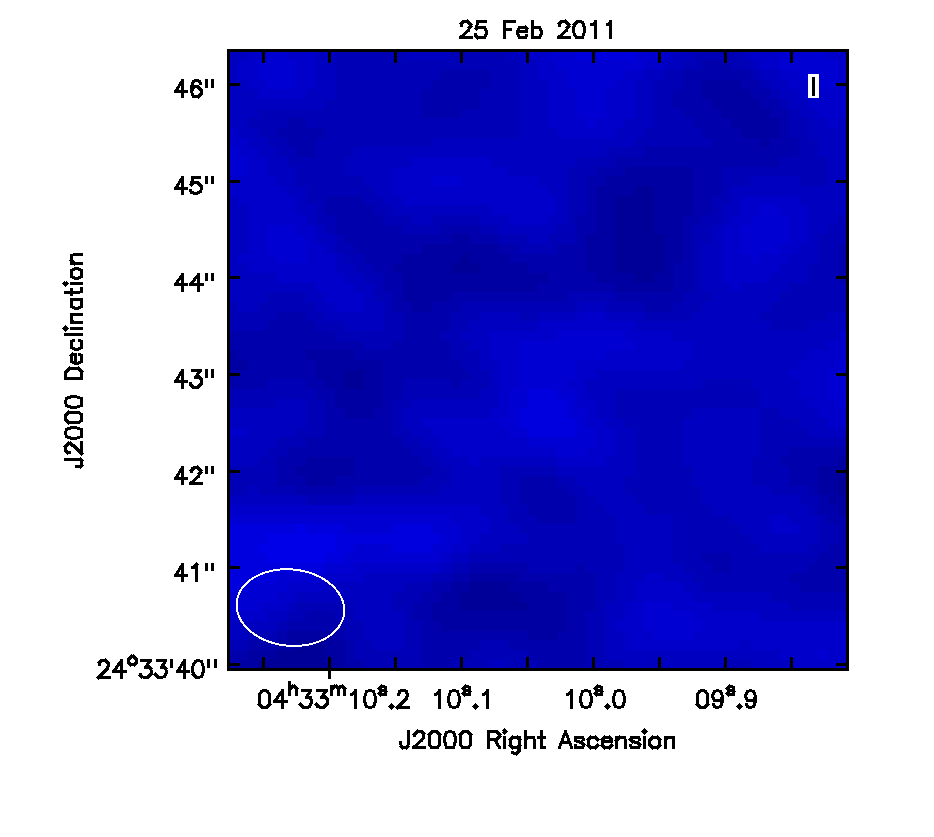}\includegraphics[width=0.33\textwidth]{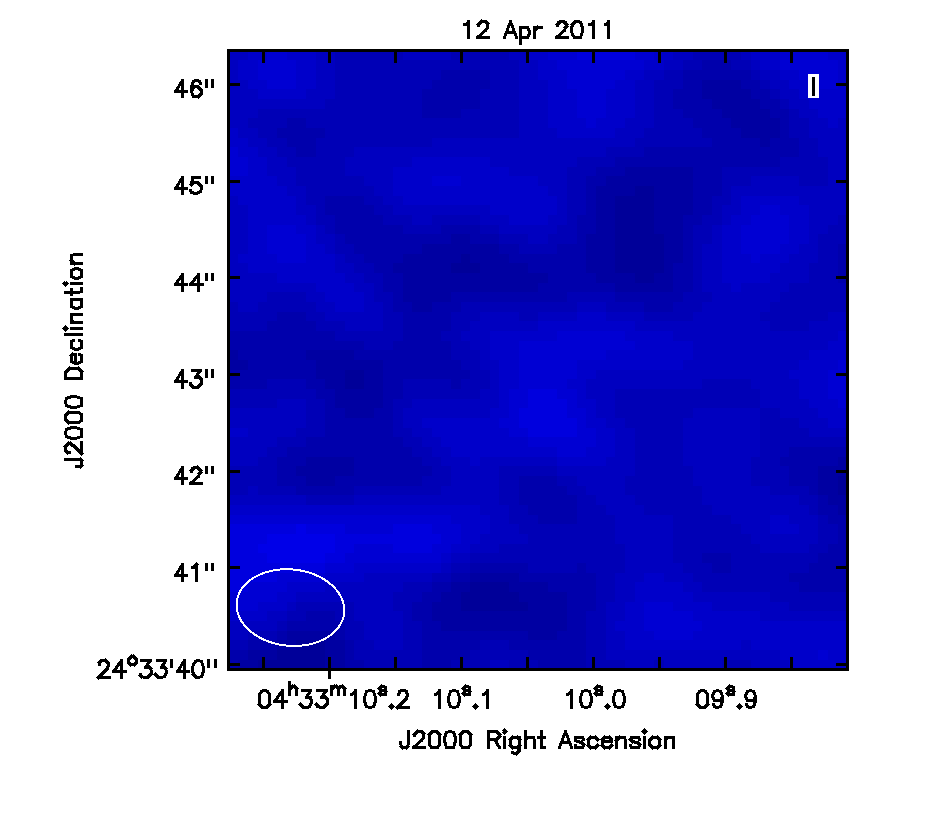}\includegraphics[width=0.33\textwidth]{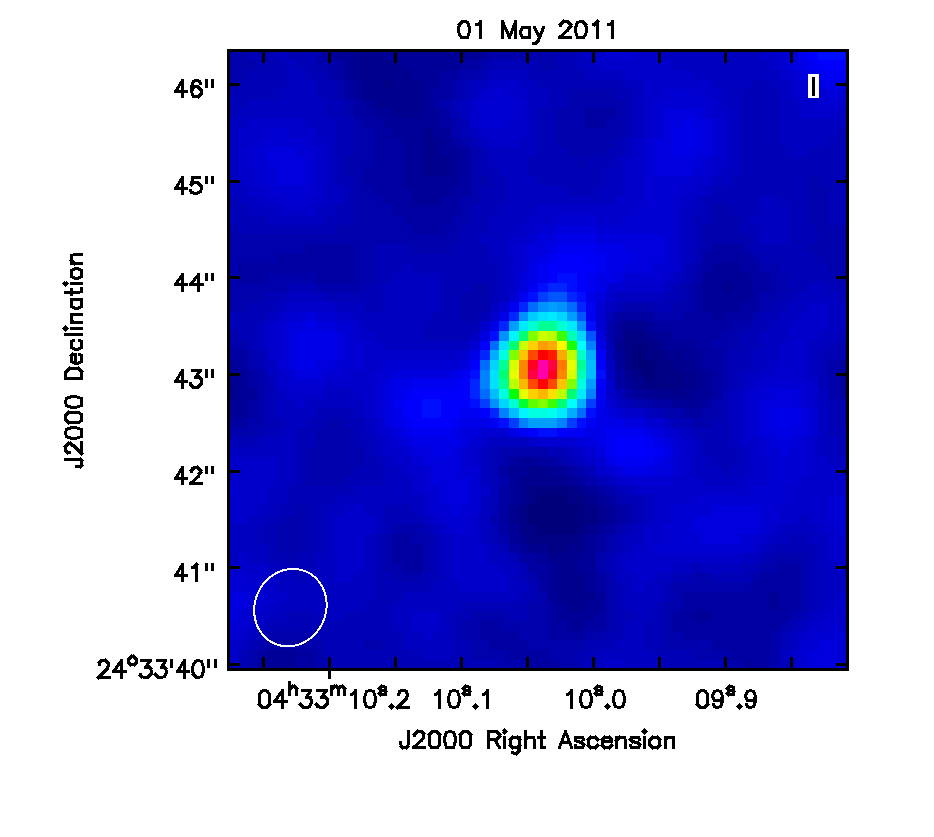}
\caption{
Images from each of the three VLA epochs, zoomed in on \vtau.  The synthesized beam is shown in the lower left.
\label{fig:v830tau}
}
\end{figure}

Figure~\ref{fig:widefield} shows the full field surrounding \vtau\ from
01 May 2011.  \vtau\ is clearly detected at the phase center of the image.
\vtau\ is not detected on 25 Feb 2011.  On 12 Apr 2011, we fit a source
at the position of \vtau\ that is nominally significant but given the presence
of other peaks of comparable brightness in the map, 
we also treat this epoch as a non-detection.
Two bright sources to the southeast of \vtau\
appear to be the dual radio lobes associated with an AGN.  
Figure~\ref{fig:v830tau} shows
narrow-field images centered on \vtau\ from all three epochs.
Table~\ref{tab:observations} summarizes the observational results.  

We also computed a Stokes V image for epoch 01 May 2011, in which
no source is detected with a $3\sigma$ upper limit of $66 \mu$Jy, or
$\sim $7\% of the peak flux density of \vtau.  Imaging \added{separately} the upper and low 
frequency windows of \vtau\ we measure a spectral index $\alpha=0.1 \pm 0.1$
for $S \propto \nu^\alpha$.  We find no statistical difference
in the flux density of \vtau\ if we split the data into two-minute
segments.

\begin{deluxetable}{llrrrr}
\tablecaption{VLA \& VLBA Results for V830 Tau \label{tab:observations}}
\tablehead{
\colhead{Tel.} & \colhead{Epoch} & \colhead{UT} & \colhead{Beam} & \colhead{RMS} & \colhead{$S$} \\
               &                   &              & 	         & \colhead{($\mu$Jy)} & \colhead{($\mu$Jy)} }
\startdata
VLA   & 25 Feb 2011 & 04:47 & $1.1" \times 0.8"$, 84$^\circ$ & 22 & $<66$  \\
\dots & 12 Apr 2011 & 02:37 & $1.6" \times 0.8"$, 86$^\circ$ & 34 & $147 \pm 34$ \\
\dots & 01 May 2011 & 22:25 & $0.8" \times 0.6"$, -23$^\circ$& 26 & $919 \pm 26$ \\
VLBA  & 31 Aug 2014 & 14:23 & $1.8 \times 0.8$ mas$^2$, -10$^\circ$ & 39 & $<117$ \\
\dots & 11 Sep 2015 & 13:42 & $1.8 \times 0.8$ mas$^2$, -11$^\circ$ & 40 & $501 \pm 75$ \\
\enddata
\tablecomments{Non-detections are given as $3\sigma$ upper limits.}
\end{deluxetable}

\subsection{Very Long Baseline Array}

V830 Tau was also observed with the Very Long Baseline Array (VLBA) on
two epochs as part of the
Gould's Belt Distances Survey (GOBELINS).
Details of the observational strategy and calibration are presented
in papers describing results for the Ophiuchus cloud from the same survey
\citep{Ortiz2016A,Ortiz2016B}. The data were obtained at a sky
frequency of 8.4 GHz with a recording bandwidth of 256 MHz in right and
left circular polarizations.  
Rapid phase switching to the compact
calibrator J0426+2327 was used to correct for short time scale atmospheric
phase fluctuations. Additionally, J0435+2532, J0429+2724 and J0438+2153 were
observed every $\sim$50 minutes to correct for the phase gradient over the sky
in the region of the target.
The data were calibrated using standard fringe-finding strategies,
achieving images with an rms flux density of $\sim 40\ \mu$Jy
and a synthesized beam of $\sim 1$ mas.
\vtau\ was detected in
only one epoch (11 September 2015) with a flux density of 501 $\pm 75\ \mu$Jy 
in a 
synthesized beam of 1.8 $\times$ 0.8 mas. The lack of detection in the other
epoch
further confirms the high variability of the source.
Results are summarized in Table~\ref{tab:observations}.

\section{Analysis and Discussion}

\vtau\ is clearly detected with the VLA on 01 May 2011 with a flux density of
$919 \pm 26$ $\mu$Jy and with the VLBA on 11 Sep 2015 with a flux density
of $501 \pm 75\ \mu$Jy.  Source fitting at the position of \vtau\ on
\deleted{on} 12 Apr 2011 reveals a marginal detection with a flux density $\sim 150$ $\mu$Jy, however, there are other features in the image with similar flux
densities, so we treat this as a non-detection.  We find
no VLA detection on 25 Feb 2011 at a $3\sigma$ threshold of
$66 \mu$Jy and no VLBA detection on 31 Aug 2014 at a $3\sigma$ threshold of 
$117\ \mu$Jy.   These non-detection are consistent
with the one reported observation in the literature, an upper limit of $110$ $\mu$Jy
at 3.6 cm from VLA observations in 1994 \citep{1996AJ....111..355C}.
Together, these results demonstrate that \vtau\ is strongly variable
on a time scale of weeks with an amplitude of variability 
greater than $10 \times$.

\vtau\ is associated with a weak X-ray source, 	1RXS J043309.8+243359,
with a luminosity $L_X \sim 10^{30}$ erg s$^{-1}$ -- $5.7 \times 10^{30}$
erg s$^{-1}$ 
\citep{1999A&A...349..389V,2007A&A...473..589S}.  The radio/X-ray 
luminosity correlation for stars, $L_X/L_R \approx 10^{15 \pm 1}$,
 has been shown to be valid over 10 orders
of magnitude and a wide range of active sources including the Sun,
M dwarfs, and young stellar objects
\citep{2002ARA&A..40..217G}.  
We estimate a radio flux density of $\sim 40$ -- 200 $\mu$Jy given
the range of X-ray measurements.  Taking into account the   
non-simultaneous nature of the radio and X-ray measurements
and the order of magnitude scatter in the correlation, we consider
our radio results to be consistent with the G\"udel-Benz
relationship.

The radio and X-ray properties of \vtau\ are consistent with those of other T Tauri stars.
Many T Tauri stars are known to be bright and variable in the radio.
For instance, GMR-A was observed to vary by more than a factor of 20$\times$
at wavelengths from 2 cm to 3mm
\citep{2003ApJ...598.1140B}.  Monitoring of the Orion, Taurus, and other regions
has identified young stellar objects with mJy flux densities,
many of them variable \citep{1987ApJ...314..535G,1993A&AS..101..127F,1997ApJ...490..735C,2012MNRAS.423.1089A,2014ApJ...790...49K,2015ApJ...801...91D}.  
X-ray and radio luminosity
and variability have been shown to be correlated in GMR-A and in other
T Tauri stars.
We do note that \citet{Ortiz2016A}
showed that T Tauri radio emission typically breaks into 
nonthermal and thermal emitters.  \vtau\ clearly falls into the 
nonthermal category.

We infer that 
the emission mechanism is gyro-synchrotron or synchrotron radiation from a power-law distribution of
relativistic particles
in a strong magnetic field based on several lines of evidence.  One, 
the VLBA detection demonstrates a nonthermal origin for emission with
the brightness temperature $T_b \gsim 1.3 \times 10^{7}$ K.
Two, the
spectral index $\alpha\approx 0$ is consistent with the partially optically thick component
of gyro-synchrotron or synchrotron radiation \citep{1985ARA&A..23..169D}.  Three, the absence of circular polarization
is consistent with the emission mechanism not originating from the cyclotron
mechanism or a coherent emission mechanism, which are known to produce
circular polarization fractions as high as 100\%  \citep{2003A&A...406..957S}.
Four, the average stellar magnetic dipole field inferred for optical
spectropolarimetry of $B \approx 350$ G leads
to a cyclotron frequency of $\sim 1$ GHz, well below the frequency of our
detections.  Variations in the field strength in the magnetosphere, however, 
could lead to regions where the cyclotron frequency is significantly higher.

Further, we can show that there is consistency between the stellar model
and gyro-synchrotron/synchrotron parameters.  \vtau\ is compact in the VLBA observations
with a source size that is equal to or smaller than the beam size $\sim 50 R_\odot$ or 0.3 AU.
Further, given the observed brightness temperature lower limit, synchrotron
radiation from a power-law distribution of electrons 
requires a magnetic field $B \gsim 30$ G \citep{2002ARA&A..40..217G},
consistent with the spectropolarimetric magnetic field average.
Therefore, we see good consistency between the
observed stellar size, magnetic field strength, and the observed radio 
emission during the high or flaring state.  In the quiescent or low state where we
do not detect \vtau, we
can only place an upper limit on $T_b < 6 \times 10^5$ K.  The detections
likely occur in a flaring
state that is the result of electron acceleration via magnetic
reconnection or other processes that produces 
a nonthermal high-energy tail for the electron energy distribution.

From these data,
we cannot infer whether the emission arises from the star, the
hot Jupiter, or an interaction between the two, although stellar
emission is the most likely explanation.  
Future observations, however, may be
able to detect the role of the hot Jupiter on the radio emission.  
We consider several possibilities.

The non-thermal radio emission from \vtau\ is compact 
and detectable with very long baseline interferometry.
High resolution imaging may be effective at 
separating stellar and exoplanet emission.  For a semimajor axis 
$a=0.057$ AU, the angular separation is 0.44 mas.  At long centimeter
wavelengths ($\sim 3$ cm), this separation will be unresolved 
but could potentially be detected with sufficient sensitivity and calibration
accuracy.
If the flat spectrum of \vtau\ persists to shorter wavelengths 
($\sim 7$ mm), then the angular separation
of the star and exoplanet
can be resolved by Earth baselines.
The astrometric reflex motion of the star
due to exoplanet, on the other hand, is equal to the angular separation
reduced by a factor $\sim M_{p}/M_\sun \approx 7.3 \times 10^{-4}$, resulting in 
an 0.35 $\mu$arcsec signal.
Astrometry on stars has achieved $\sim 100\ \mu$arcsec accuracy but
without detection of an exoplanet companion, although numerous
stellar binaries have been characterized
\citep[e.g.,][]{2009ApJ...701.1922B,2011ApJ...740...32B,2013ApJ...777...70F,2014ARA&A..52..339R,2016arXiv160507177F,Ortiz2016B}.  If the presence of a hot Jupiter is
indicative of a more complex planetary system with Jupiter mass
planets at large semi-major axes, then astrometric detection of that
signature may be possible.

\begin{figure}
\includegraphics[width=\textwidth]{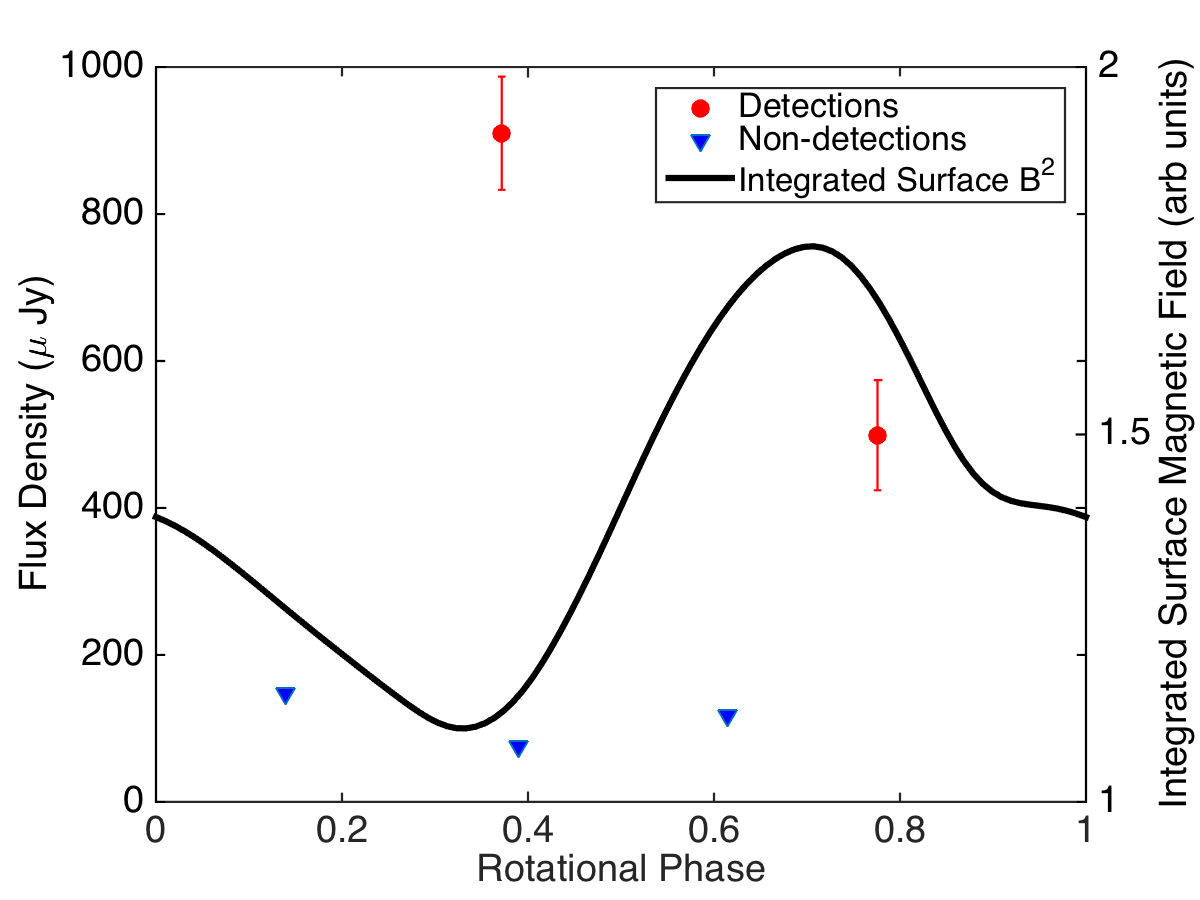}
\caption{Radio flux density as a function of stellar rotational phase.
The left y-axis shows the range of flux densities, which are plotted
for VLA and VLBA detections and non-detections ($3\sigma$ upper limits).
The right y-axis shows the relative scaling of the integrated surface
magnetic field energy, which was determined from 2015 observations
and is plotted as the dark line.  Differential rotation and changes 
in the stellar dynamo imply that only radio
observations obtained within $\sim 1$ year of the spectropolarimetric
data are likely to be meaningfully related to the shape 
of the magnetic energy curve.  Thus, only the VLBA detection at phase of
0.8 has a direct relationship with the magnetic energy curve.
\label{fig:rotphase}
}
\end{figure}

Total intensity monitoring across the radio spectrum
offers promising opportunities for characterizing the stellar and
exoplanet magnetosphere.  We plot in Figure~\ref{fig:rotphase},
the detected flux density as a function of stellar rotational phase, using
the ephemeris from \citet{2016Natur.534..662D}.  There is no
clear trend in these sparse data with rotational phase.  This is likely
the result of the detections being the result of flaring activity in \vtau,
but may also be due to variable magnetospheric structure.  
If it is detectable,
quiescent or steady radio emission is more likely than flaring flux
to show a trend with
rotational phase.  We also plot the mean
surface magnetic field density as a function of rotational phase 
in Figure~\ref{fig:rotphase} based on the Dec 2015 observations.  This
calculation approximates the total volume-integrated magnetic energy density,
albeit with an emphasis on the surface field.  Synchrotron radiation
is proportional to the magnetic energy density, and, therefore, in the
case of a uniform distribution of nonthermal particles will
follow this magnetic quantity.  The total variation in the computed
quantity is a factor of 1.6, far less than the factor of $>10\times$ variability
observed.  Differential rotation and other secular
effects on the magnetosphere imply that the particular profile of the
magnetic field is likely to evolve on a time scale of the differential
rotation $\sim 1$ yr.
This makes clear that simultaneous and/or contemporaneous radio and spectropolarimetric is
necessary to fully disentangle any effects associated with the rotational
period.

We cannot make a meaningful
comparison of the existing radio data with orbital phase, because the
latter is only known to an accuracy of 1\%.  Over the $\gsim 4$ year span
of our observations, we completely lose any orbital phase information.
Currently, the eccentricity of the orbit is weakly constrained to
$e < 0.3$.  Highly eccentric orbits in stellar binaries have been shown to
produce periodic radio emission \citep{2002A&A...382..152M,2011ApJ...743..175A,2012ApJ...747...18T}. 
Dense sampling with sensitivity to quiescent flux 
over multiple beat periods ($P_{beat}=6.17$ d)
between the orbital and rotational periods 
will be necessary to discriminate between rotational and orbital effects
\citep{2010MNRAS.406..409F}.
Observations over even longer time intervals 
are also necessary to
characterize the time scale and nature of variability.

At meter wavelengths, stellar and exoplanet emission may become substantially 
brighter.  For the star, the cyclotron frequency is 1 GHz for a mean
magnetic field of 350 G, indicating the possibility for strong
low frequency emission.  The magnetic field of the exoplanet, of course, is unknown.  But it
is not unreasonable to think that the magnetic field for a young hot Jupiter would
be higher than that of an  older one, just as the young stellar magnetic field is 
higher.  \replaced{Following the}{The} radiometric Bode's law
\citep{2001Ap&SS.277..293Z,2004ApJ...612..511L} 
\added{provides a method for estimating the coherent cyclotron 
radio luminosity of a star-exoplanet system based on scalings
observed in the solar system.  We use the law to predict}
the median 
radio luminosity for a hot Jupiter orbiting a solar-type star 
\replaced{is}{to be} $P=3 \times 10^{21}$ erg s$^{-1}$, corresponding to a flux density of 
$\sim 1$ $\mu$Jy for \vtau\ at a frequency of 100 MHz.
The radio flux, however, scales with the stellar wind power, likely proportional
to the stellar magnetic field energy density, and with 
the planetary magnetic moment, which is also likely
proportional to the stellar magnetic field strength.  Thus, the radio power may
scale as $\sim B^3$, which is $10^3$ to $10^6$
higher for \vtau\ than for the Sun.  Thus, a peak flux density of $\sim 1$ --- 1000 mJy from the exoplanet at meter wavelengths is feasible for \vtau.  
\added{Estimates of this kind are, of course, very uncertain given
the absence of any exoplanet detections at radio wavelengths.}
GMRT observations at 323 and 608 MHz of other young stellar objects in
Taurus indicate mJy flux levels that are consistent with free-free
emission \added{rather than a coherent emission process}
\citep{2016MNRAS.459.1248A}.

The discovery of variable radio emission from \vtau\ 
\replaced{opens a new window
for the discovery and characterization of exoplanets.}{opens a new observational window on hot jupiter hosts and, potentially, on the
star-planet magnetic interaction.}
For the first time, we have detected the stellar
host of an exoplanet system.  Radio observations can play an important role in full
characterization of the stellar magnetosphere and may ultimately be able to detect
the exopolanet  or star-exoplanet interactions, either directly or indirectly.  Simultaneous observations over multiple cycles of the rotational and orbital
periods are essential for accurate characterization \added{and exploration
of this unique system}.  
The study of radio
emission from other T Tauri stars can provide a statistical characterization of the 
magnetic and particle environments of young stars \added{and, possibly, their
planetary companions}.

\acknowledgements

The National Radio Astronomy Observatory is a facility of the National Science Foundation operated under cooperative agreement by Associated Universities, Inc. 
L.L., S.A.D.Q, and G.N.O.-L acknowledge the financial support of DGAPA, UNAM, and CONACyT, M\'exico. L.L. acknowledges financial support from the von Humboldt Stiftung.
P.A.B.G. acknowledges financial support from FAPESP.
J.F.D. thanks the IDEX initiative of Universit\'e F\'ed\'erale Toulouse 
Midi-Pyr\'en\'ees for awarding a Chaire d’Attractivit\'e in the framework of which the study of \vtau\ was carried out.


\begin{thebibliography}{}
\expandafter\ifx\csname natexlab\endcsname\relax\def\natexlab#1{#1}\fi

\bibitem[{{Abrevaya} {et~al.}(2012){Abrevaya}, {Cort{\'o}n}, \&
  {Mauas}}]{2012IAUS..286..405A}
{Abrevaya}, X.~C., {Cort{\'o}n}, E., \& {Mauas}, P.~J.~D. 2012, in IAU
  Symposium, Vol. 286, Comparative Magnetic Minima: Characterizing Quiet Times
  in the Sun and Stars, ed. C.~H. {Mandrini} \& D.~F. {Webb}, 405--409

\bibitem[{{Adams} {et~al.}(2011){Adams}, {Cai}, {Galli}, {Lizano}, \&
  {Shu}}]{2011ApJ...743..175A}
{Adams}, F.~C., {Cai}, M.~J., {Galli}, D., {Lizano}, S., \& {Shu}, F.~H. 2011,
  \apj, 743, 175

\bibitem[{{Ainsworth} {et~al.}(2016){Ainsworth}, {Scaife}, {Green}, {Coughlan},
  \& {Ray}}]{2016MNRAS.459.1248A}
{Ainsworth}, R.~E., {Scaife}, A.~M.~M., {Green}, D.~A., {Coughlan}, C.~P., \&
  {Ray}, T.~P. 2016, \mnras, 459, 1248

\bibitem[{{AMI Consortium} {et~al.}(2012){AMI Consortium}, {Ainsworth},
  {Scaife}, {Ray}, {Buckle}, {Davies}, {Franzen}, {Grainge}, {Hobson},
  {Hurley-Walker}, {Lasenby}, {Olamaie}, {Perrott}, {Pooley}, {Richer},
  {Rodr{\'{\i}}guez-Gonz{\'a}lvez}, {Saunders}, {Schammel}, {Scott},
  {Shimwell}, {Titterington}, \& {Waldram}}]{2012MNRAS.423.1089A}
{AMI Consortium}, {Ainsworth}, R.~E., {Scaife}, A.~M.~M., {et~al.} 2012,
  \mnras, 423, 1089

\bibitem[{{Bower} {et~al.}(2011){Bower}, {Bolatto}, {Ford}, {Fries}, {Kalas},
  {Sanchez}, {Sanderbeck}, \& {Viscomi}}]{2011ApJ...740...32B}
{Bower}, G.~C., {Bolatto}, A., {Ford}, E.~B., {et~al.} 2011, \apj, 740, 32

\bibitem[{{Bower} {et~al.}(2009){Bower}, {Bolatto}, {Ford}, \&
  {Kalas}}]{2009ApJ...701.1922B}
{Bower}, G.~C., {Bolatto}, A., {Ford}, E.~B., \& {Kalas}, P. 2009, \apj, 701,
  1922

\bibitem[{{Bower} {et~al.}(2003){Bower}, {Plambeck}, {Bolatto}, {McCrady},
  {Graham}, {de Pater}, {Liu}, \& {Baganoff}}]{2003ApJ...598.1140B}
{Bower}, G.~C., {Plambeck}, R.~L., {Bolatto}, A., {et~al.} 2003, \apj, 598,
  1140

\bibitem[{{Carkner} {et~al.}(1997){Carkner}, {Mamajek}, {Feigelson},
  {Neuh{\"a}user}, {Wichmann}, \& {Krautter}}]{1997ApJ...490..735C}
{Carkner}, L., {Mamajek}, E., {Feigelson}, E., {et~al.} 1997, \apj, 490, 735

\bibitem[{{Chene} {et~al.}(2014){Chene}, {Padzer}, {Barrick}, {Anthony},
  {Benedict}, {Duncan}, {Gigoux}, {Kleinman}, {Malo}, {Martioli}, {Moutou},
  {Placco}, {Reshetovand}, {Rhee}, {Roth}, {Schiavon}, {Tollestrup},
  {Vermeulen}, {White}, \& {Wooff}}]{2014SPIE.9151E..47C}
{Chene}, A.-N., {Padzer}, J., {Barrick}, G., {et~al.} 2014, in \procspie, Vol.
  9151, Advances in Optical and Mechanical Technologies for Telescopes and
  Instrumentation, 915147

\bibitem[{{Chiang} {et~al.}(1996){Chiang}, {Phillips}, \&
  {Lonsdale}}]{1996AJ....111..355C}
{Chiang}, E., {Phillips}, R.~B., \& {Lonsdale}, C.~J. 1996, \aj, 111, 355

\bibitem[{{Donati}(2003)}]{2003ASPC..307...41D}
{Donati}, J.-F. 2003, in Astronomical Society of the Pacific Conference Series,
  Vol. 307, Solar Polarization, ed. J.~{Trujillo-Bueno} \& J.~{Sanchez
  Almeida}, 41

\bibitem[{{Donati} {et~al.}(2015){Donati}, {H{\'e}brard}, {Hussain}, {Moutou},
  {Malo}, {Grankin}, {Vidotto}, {Alencar}, {Gregory}, {Jardine}, {Herczeg},
  {Morin}, {Fares}, {M{\'e}nard}, {Bouvier}, {Delfosse}, {Doyon}, {Takami},
  {Figueira}, {Petit}, {Boisse}, \& {MaTYSSE
  Collaboration}}]{2015MNRAS.453.3706D}
{Donati}, J.-F., {H{\'e}brard}, E., {Hussain}, G.~A.~J., {et~al.} 2015, \mnras,
  453, 3706

\bibitem[{{Donati} {et~al.}(2016){Donati}, {Moutou}, {Malo}, {Baruteau}, {Yu},
  {H{\'e}brard}, {Hussain}, {Alencar}, {M{\'e}nard}, {Bouvier}, {Petit},
  {Takami}, {Doyon}, \& {Cameron}}]{2016Natur.534..662D}
{Donati}, J.~F., {Moutou}, C., {Malo}, L., {et~al.} 2016, \nat, 534, 662

\bibitem[{{Dulk}(1985)}]{1985ARA&A..23..169D}
{Dulk}, G.~A. 1985, \araa, 23, 169

\bibitem[{{Dzib} {et~al.}(2015){Dzib}, {Loinard}, {Rodr{\'{\i}}guez},
  {Mioduszewski}, {Ortiz-Le{\'o}n}, {Kounkel}, {Pech}, {Rivera}, {Torres},
  {Boden}, {Hartmann}, {Evans}, {Brice{\~n}o}, \&
  {Tobin}}]{2015ApJ...801...91D}
{Dzib}, S.~A., {Loinard}, L., {Rodr{\'{\i}}guez}, L.~F., {et~al.} 2015, \apj,
  801, 91

\bibitem[{{Fares} {et~al.}(2010){Fares}, {Donati}, {Moutou}, {Jardine},
  {Grie{\ss}meier}, {Zarka}, {Shkolnik}, {Bohlender}, {Catala}, \& {Collier
  Cameron}}]{2010MNRAS.406..409F}
{Fares}, R., {Donati}, J.-F., {Moutou}, C., {et~al.} 2010, \mnras, 406, 409

\bibitem[{{Farrell} {et~al.}(2003){Farrell}, {Desch}, {Lazio}, {Bastian}, \&
  {Zarka}}]{2003ASPC..294..151F}
{Farrell}, W.~M., {Desch}, M.~D., {Lazio}, T.~J., {Bastian}, T., \& {Zarka}, P.
  2003, in Astronomical Society of the Pacific Conference Series, Vol. 294,
  Scientific Frontiers in Research on Extrasolar Planets, ed. D.~{Deming} \&
  S.~{Seager}, 151--156

\bibitem[{{Farrell} {et~al.}(2004){Farrell}, {Lazio}, {Desch}, {Bastian}, \&
  {Zarka}}]{2004IAUS..213...73F}
{Farrell}, W.~M., {Lazio}, T.~J.~W., {Desch}, M.~D., {Bastian}, T.~S., \&
  {Zarka}, P. 2004, in IAU Symposium, Vol. 213, Bioastronomy 2002: Life Among
  the Stars, ed. R.~{Norris} \& F.~{Stootman}, 73

\bibitem[{{Feigelson} {et~al.}(2002){Feigelson}, {Broos}, {Gaffney}, {Garmire},
  {Hillenbrand}, {Pravdo}, {Townsley}, \& {Tsuboi}}]{2002ApJ...574..258F}
{Feigelson}, E.~D., {Broos}, P., {Gaffney}, III, J.~A., {et~al.} 2002, \apj,
  574, 258

\bibitem[{{Feigelson} \& {Montmerle}(1999)}]{1999ARA&A..37..363F}
{Feigelson}, E.~D., \& {Montmerle}, T. 1999, \araa, 37, 363

\bibitem[{{Felli} {et~al.}(1993){Felli}, {Taylor}, {Catarzi}, {Churchwell}, \&
  {Kurtz}}]{1993A&AS..101..127F}
{Felli}, M., {Taylor}, G.~B., {Catarzi}, M., {Churchwell}, E., \& {Kurtz}, S.
  1993, \aaps, 101, 127

\bibitem[{{Forbrich} {et~al.}(2013){Forbrich}, {Berger}, \&
  {Reid}}]{2013ApJ...777...70F}
{Forbrich}, J., {Berger}, E., \& {Reid}, M.~J. 2013, \apj, 777, 70

\bibitem[{{Forbrich} {et~al.}(2016){Forbrich}, {Dupuy}, {Reid}, {Berger},
  {Rizzuto}, {Mann}, {Liu}, {Aller}, \& {Kraus}}]{2016arXiv160507177F}
{Forbrich}, J., {Dupuy}, T.~J., {Reid}, M.~J., {et~al.} 2016, ArXiv e-prints,
  arXiv:1605.07177

\bibitem[{{Garay} {et~al.}(1987){Garay}, {Moran}, \&
  {Reid}}]{1987ApJ...314..535G}
{Garay}, G., {Moran}, J.~M., \& {Reid}, M.~J. 1987, \apj, 314, 535

\bibitem[{{Grankin}(2013)}]{2013AstL...39..251G}
{Grankin}, K.~N. 2013, Astronomy Letters, 39, 251

\bibitem[{{G{\"u}del}(2002)}]{2002ARA&A..40..217G}
{G{\"u}del}, M. 2002, \araa, 40, 217

\bibitem[{{Hallinan} {et~al.}(2013){Hallinan}, {Sirothia}, {Antonova},
  {Ishwara-Chandra}, {Bourke}, {Doyle}, {Hartman}, \&
  {Golden}}]{2013ApJ...762...34H}
{Hallinan}, G., {Sirothia}, S.~K., {Antonova}, A., {et~al.} 2013, \apj, 762, 34

\bibitem[{{Kounkel} {et~al.}(2014){Kounkel}, {Hartmann}, {Loinard},
  {Mioduszewski}, {Dzib}, {Ortiz-Le{\'o}n}, {Rodr{\'{\i}}guez}, {Pech},
  {Rivera}, {Torres}, {Boden}, {Evans}, {Brice{\~n}o}, \&
  {Tobin}}]{2014ApJ...790...49K}
{Kounkel}, M., {Hartmann}, L., {Loinard}, L., {et~al.} 2014, \apj, 790, 49

\bibitem[{{Lazio} {et~al.}(2004){Lazio}, {Farrell}, {Dietrick}, {Greenlees},
  {Hogan}, {Jones}, \& {Hennig}}]{2004ApJ...612..511L}
{Lazio}, W., T.~J., {Farrell}, W.~M., {Dietrick}, J., {et~al.} 2004, \apj, 612,
  511

\bibitem[{{Lazio} \& {Farrell}(2007)}]{2007ApJ...668.1182L}
{Lazio}, T.~J.~W., \& {Farrell}, W.~M. 2007, \apj, 668, 1182

\bibitem[{{Lazio} {et~al.}(2010){Lazio}, {Shankland}, {Farrell}, \&
  {Blank}}]{2010AJ....140.1929L}
{Lazio}, T.~J.~W., {Shankland}, P.~D., {Farrell}, W.~M., \& {Blank}, D.~L.
  2010, \aj, 140, 1929

\bibitem[{{Lecavelier Des Etangs} {et~al.}(2009){Lecavelier Des Etangs},
  {Sirothia}, {Gopal-Krishna}, \& {Zarka}}]{2009A&A...500L..51L}
{Lecavelier Des Etangs}, A., {Sirothia}, S.~K., {Gopal-Krishna}, \& {Zarka}, P.
  2009, \aap, 500, L51

\bibitem[{{Massi} {et~al.}(2002){Massi}, {Menten}, \&
  {Neidh{\"o}fer}}]{2002A&A...382..152M}
{Massi}, M., {Menten}, K., \& {Neidh{\"o}fer}, J. 2002, \aap, 382, 152

\bibitem[{{Ortiz-Le{\'o}n} {et~al.}(2016{\natexlab{a}}){Ortiz-Le{\'o}n},
  {Loinard}, {Dzib}, {Mioduszewski}, {Rodriguez}, {Torres},
  {Gonzalez-Lopezlira}, {Pech}, {Rivera}, {Kounkel}, {Hartmann}, {Boden},
  {Evans II}, {Briceno}, {Tobin}, \& {Galli}}]{Ortiz2016A}
{Ortiz-Le{\'o}n}, G.~N., {Loinard}, L., {Dzib}, S.~A., {et~al.}
  2016{\natexlab{a}}, \apj, submitted

\bibitem[{{Ortiz-Le{\'o}n} {et~al.}(2016{\natexlab{b}}){Ortiz-Le{\'o}n},
  {Loinard}, {Kounkel}, {Dzib}, {Mioduszewski}, {Rodriguez}, {Torres},
  {Gonzalez-Lopezlira}, {Pech}, {Rivera}, , {Hartmann}, {Boden}, {Evans II},
  {Briceno}, {Tobin}, {Galli}, \& {Gudehus}}]{Ortiz2016B}
{Ortiz-Le{\'o}n}, G.~N., {Loinard}, L., {Kounkel}, M.~A., {et~al.}
  2016{\natexlab{b}}, \apj, submitted

\bibitem[{{Osten} \& {Wolk}(2015)}]{2015ApJ...809...79O}
{Osten}, R.~A., \& {Wolk}, S.~J. 2015, \apj, 809, 79

\bibitem[{{Reid} \& {Honma}(2014)}]{2014ARA&A..52..339R}
{Reid}, M.~J., \& {Honma}, M. 2014, \araa, 52, 339

\bibitem[{{Scelsi} {et~al.}(2007){Scelsi}, {Maggio}, {Micela}, {Briggs}, \&
  {G{\"u}del}}]{2007A&A...473..589S}
{Scelsi}, L., {Maggio}, A., {Micela}, G., {Briggs}, K., \& {G{\"u}del}, M.
  2007, \aap, 473, 589

\bibitem[{{Siess} {et~al.}(2000){Siess}, {Dufour}, \&
  {Forestini}}]{2000A&A...358..593S}
{Siess}, L., {Dufour}, E., \& {Forestini}, M. 2000, \aap, 358, 593

\bibitem[{{Sirothia} {et~al.}(2014){Sirothia}, {Lecavelier des Etangs},
  {Gopal-Krishna}, {Kantharia}, \& {Ishwar-Chandra}}]{2014A&A...562A.108S}
{Sirothia}, S.~K., {Lecavelier des Etangs}, A., {Gopal-Krishna}, {Kantharia},
  N.~G., \& {Ishwar-Chandra}, C.~H. 2014, \aap, 562, A108

\bibitem[{{Smith} {et~al.}(2003){Smith}, {Pestalozzi}, {G{\"u}del}, {Conway},
  \& {Benz}}]{2003A&A...406..957S}
{Smith}, K., {Pestalozzi}, M., {G{\"u}del}, M., {Conway}, J., \& {Benz}, A.~O.
  2003, \aap, 406, 957

\bibitem[{{Torres} {et~al.}(2012){Torres}, {Loinard}, {Mioduszewski}, {Boden},
  {Franco-Hern{\'a}ndez}, {Vlemmings}, \&
  {Rodr{\'{\i}}guez}}]{2012ApJ...747...18T}
{Torres}, R.~M., {Loinard}, L., {Mioduszewski}, A.~J., {et~al.} 2012, \apj,
  747, 18

\bibitem[{{Torres} {et~al.}(2009){Torres}, {Loinard}, {Mioduszewski}, \&
  {Rodr{\'{\i}}guez}}]{2009ApJ...698..242T}
{Torres}, R.~M., {Loinard}, L., {Mioduszewski}, A.~J., \& {Rodr{\'{\i}}guez},
  L.~F. 2009, \apj, 698, 242

\bibitem[{{Voges} {et~al.}(1999){Voges}, {Aschenbach}, {Boller},
  {Br{\"a}uninger}, {Briel}, {Burkert}, {Dennerl}, {Englhauser}, {Gruber},
  {Haberl}, {Hartner}, {Hasinger}, {K{\"u}rster}, {Pfeffermann}, {Pietsch},
  {Predehl}, {Rosso}, {Schmitt}, {Tr{\"u}mper}, \&
  {Zimmermann}}]{1999A&A...349..389V}
{Voges}, W., {Aschenbach}, B., {Boller}, T., {et~al.} 1999, \aap, 349, 389

\bibitem[{{Winterhalter} {et~al.}(2015){Winterhalter}, {Knapp}, {Majid},
  {Lazio}, {Farrell}, \& {Splitter}}]{2015EGUGA..1714693W}
{Winterhalter}, D., {Knapp}, M., {Majid}, W., {et~al.} 2015, in EGU General
  Assembly Conference Abstracts, Vol.~17, EGU General Assembly Conference
  Abstracts, 14693

\bibitem[{{Wolszczan} \& {Frail}(1992)}]{1992Natur.355..145W}
{Wolszczan}, A., \& {Frail}, D.~A. 1992, \nat, 355, 145

\bibitem[{{Zarka} {et~al.}(2015){Zarka}, {Lazio}, \&
  {Hallinan}}]{2015aska.confE.120Z}
{Zarka}, P., {Lazio}, J., \& {Hallinan}, G. 2015, Advancing Astrophysics with
  the Square Kilometre Array (AASKA14), 120

\bibitem[{{Zarka} {et~al.}(2001){Zarka}, {Treumann}, {Ryabov}, \&
  {Ryabov}}]{2001Ap&SS.277..293Z}
{Zarka}, P., {Treumann}, R.~A., {Ryabov}, B.~P., \& {Ryabov}, V.~B. 2001,
  \apss, 277, 293

\end{thebibliography}

\listofchanges
\end{document}